# Abundance, Sizes, and Major Element Compositions of Components in CR and LL Chondrites: Formation from Single Reservoirs

Denton S. EBEL[1,2,3*], Marina E. GEMMA[1,4], Samuel P. ALPERT[1,3], Jasmine BAYRON[5], Ana H. LOBO[6], and Michael K. WEISBERG[7,3,1]

[1]Department of Earth and Planetary Sciences, American Museum of Natural History, New York, NY, 10024, USA,
[2]Lamont-Doherty Earth Observatory, Columbia University, Palisades, New York, 10964, USA,
[3]Department of Earth and Environmental Sciences, Graduate Center of the City University of New York, New York, NY, 10065,
[4]Department of Geosciences, Stony Brook University, Stony Brook, NY, 11794, USA
[5]Department of Geography, Hunter College, City University of New York, New York, NY, 10065, USA
[6]Department Physics & Astronomy, University of California Irvine, Irvine, CA, 92697, USA,
[7]Department of Physical Sciences, Kingsborough College, City University of New York, Brooklyn, NY, 11235, USA

* Corresponding author: debel@amnh.org
ORCID numbers:
Denton Ebel: 0000-0003-2219-4902
Marina Gemma: 0000-0002-8871-773X
Samuel Alpert: 0000-0002-1513-8091
Jasmine Bayron: 0000-0001-8240-046X
Ana Lobo: 0000-0003-3862-1817
Michael Weisberg: 0000-0003-1918-8026

*Meteoritics & Planetary Sciences* (20pp)
**https://doi.org/10.1111/maps.14191**

Submitted 29-February-2024
Revised 29-April-2024
Accepted 5-June-2024
Published online 11-June-2024

extensive supplemental data is deposited online

**Abstract** --

Abundances, apparent sizes, and individual chemical compositions of chondrules, refractory inclusions, other objects and surrounding matrix have been determined for Semarkona (LL3.00) and Renazzo (CR2) using consistent methods and criteria on x-ray element intensity maps. These represent the non-carbonaceous (NC, Semarkona) and carbonaceous chondrite (CC, Renazzo) superclans of chondrite types. We compare object and matrix abundances with similar data for CM, CO, K, and CV chondrites. We assess, pixel-by-pixel, the major element abundance in each object and in the entire matrix. We determine the abundance of "metallic chondrules" in LL chondrites. Chondrules with high Mg/Si and low Fe/Si and matrix carrying opposing ratios complement each other to make whole rocks with near-solar major element ratios in Renazzo. Similar Mg/Si and Fe/Si chondrule-matrix relationships are seen in Semarkona, which is within 11% of solar Mg/Si but significantly Fe-depleted. These results provide a robust constraint in support of single-reservoir models for chondrule formation and accretion, ruling out whole classes of astrophysical models and constraining processes of chondrite component formation and accretion into chondrite parent bodies.

## 1. INTRODUCTION

Chondritic meteorites are mixtures of components. Macroscopic components include ferro-magnesian chondrules, Ca-, Al-rich inclusions (CAIs), amoeboid olivine aggregates (AOAs), opaque assemblages (OAs), and other inclusions. We refer to all these non-matrix objects as "clasts", in these sediment-like rocks. All these inclusions were once free-floating in space and show partial or complete melting at temperatures (T) above 1500 K (Ebel, 2006). The inclusions are surrounded by a fine-grained matrix that includes presolar grains formed around other stars, organic matter, and amorphous silicates, none of which could survive T above ~600 K. Matrix abundance varies from 99% to <25% among chondrites (Weisberg et al., 2006). Chondrules dominate the object (clast) population in all chondrites except the nearly clast-free CI group (Frank et al., 2023).

All chondritic meteorite evidence is overprinted by parent body processes to varying degrees. The parent bodies of the meteorites were aqueously or thermally metamorphosed when they were young and warm due to radiogenic and collisional heat, and perhaps wet due to accretion of now-vanished ices. Meteorite petrology requires seeing through these processes to decipher the initial conditions of inclusion formation and accretion with matrix. Even the nebular or asteroidal origin of some fraction of the objects, such as magnetite rims on OAs (Krot et al., 2022; Alpert et al., 2021), remains in dispute. However, the least volatile major elements and refractory trace elements largely escaped remobilization by such processes (Stracke et al., 2012). Here, we focus on least altered (lowest petrologic grade) CR and LL chondrite falls (Weisberg et al., 2006).

The bulk compositions of chondritic meteorites are a fundamental constraint on their origin. Here, we focus on Si, Mg, and Fe (Lodders, 2003). The marked similarity of ordinary (OC) and carbonaceous (CC) chondrite bulk compositions in Mg, Si, and refractory lithophile elements is a first order observation (Prior, 1913; Wood, 1963). The subtle chemical variations in major non-volatile elements Si, Mg, Al, Ti, Ca and trace refractory elements (rare earth – REE, and high field strength – HFSE elements) between chondrite groups are thought to represent localized heterogeneity of the nebular environments in which chondrites formed (Grossman, 1996; Brearley and Jones, 1998; Jones and Schilk, 2009). Chondrite compositions are within a factor of two of the solar photosphere abundances of these elements and in most cases much closer; and, of course, the CI chondrite composition is the canonical "chondritic" reference (Figure 1; Lodders, 2021). The initial condition resulting in all the chondrite parent bodies is, therefore, a mixture of material with near-solar non-volatile elemental composition with the exception of Fe.

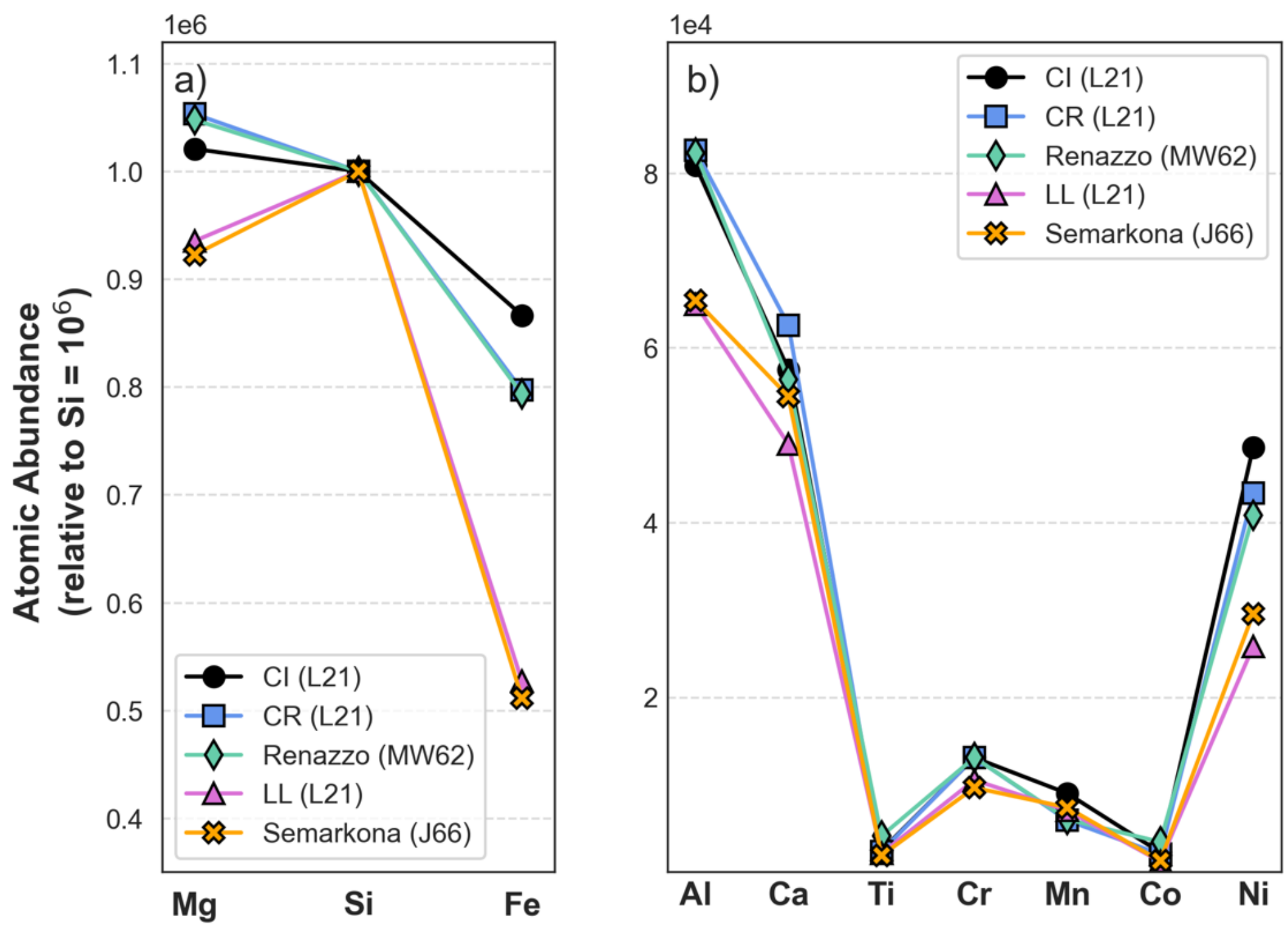

**Figure 1**: Abundances of a) major and b) minor elements in CI, CR, and LL chondrites (atoms, relative to Si = $10^6$) compiled by Lodders (2021; L21) based on all prior analyses, compared to wet chemical analyses of Renazzo (Mason & Wiik, 1962; MW62) and Semarkona (Jarosewich, 1966; J66).

Chondrites are classified into groups that each have a relatively narrow compositional range (Wasson and Kallemeyn, 1988) and contain remarkably characteristic amounts of matrix, ratios among inclusion subtypes, inclusion size distributions, and abundances of textural types of inclusions (Wood, 1963; MacPherson et al., 2005; Jones, 2012). These gross textural features are readily recognizable by meteorite petrographers using the petrographic microscope. These features initially informed the primary classification of chondrites into groups (Van Schmus and Wood, 1967; Van Schmus, 1969). These petrological observables are consistent with more recent classification based on oxygen isotopic analysis (Clayton et al., 1976; Weisberg et al., 2006).

Many current hypotheses for the formation of chondrites and their inclusions hold that refractory inclusions, chondrules, and matrix originated in different nebular regions. Most inclusions (e.g., chondrules) encountered temperatures above 1500 K (Stolper & Paque, 1986; Hewins et al., 2005). Their age, refractory nature, high $^{16}O$ content, and other isotopic observations (McKeegan et al., 2000; Kööp et al., 2018) have led to the widely held view that refractory inclusions (CAIs and AOAs) formed in the hot, innermost protoplanetary disk (e.g., Kita et al., 2013), and that chondrules formed in inner solar system regions in which matrix would not survive or in indeterminate locations with episodic heating events (Libourel and Portail, 2018). Inclusions formed in these hot regions were subsequently transported to cold regions that preserved heat-sensitive matrix components such as presolar grains (Mendybaev et al., 2002; Huss et al., 2006) and macromolecular carbon (Alexander et al., 2007). These components then accreted into the parent bodies of chondrites. The Kuiper belt comet Wild 2 also contains chondrule-like (Nakamura et al., 2008), and CAI-like high-temperature inclusions (Zolensky et al., 2006; Simon et al., 2008) interpreted as requiring transport from near the Sun (Brownlee, 2014). In this scenario, high-temperature materials were transported over large distances from multiple formation reservoirs prior to their accretion into chondrite parent bodies.

An alternative to these prevailing multiple-reservoir models is the "single reservoir" concept. This holds that temporally and/or spatially separated regions of the disk contained precursor materials from which chondrules formed by a so far unidentified localized heating process that did not destroy the surviving matrix. The inclusion-forming process operated slightly differently in each region, resulting in distinct chondrule and CAI textural populations, as is observed (Wood, 1963; Jones, 2012; MacPherson et al., 2005). Within each region, clasts and surviving matrix accreted into chondrites on timescales shorter than appreciable mixing of the chondrules with the rest of the disk. Each chondrite group thus apparently sampled a distinct population of inclusions, yet each retains near-solar ratios of most non-volatile elements. Although the CAIs are volumetrically insignificant, they are important carriers of refractory trace elements so their abundances are important to understand. Isotopic evidence supports single reservoirs of formation for chondrules and matrix in the CV group (Becker et al., 2015; Budde et al., 2016). This complementary nature of chondrules and matrix has been called "complementarity" (Palme et al., 1992; Klerner, 2001; Bland et al. 2005; Hezel and Palme 2008, 2010; Palme et al., 2014; Hezel et al., 2018; the idea can be traced to Wood, 1963, 2005). This powerful concept rules out separate reservoirs of chondrule formation and thus many models for chondrule formation.

The "single reservoir" model does not propose any particular astrophysical mechanism for heating small dust-bearing regions of the disk. It simply implies that such a mechanism must exist and operate slightly differently, or not at all, in each potential chondrite-forming reservoir. Differences between reservoirs would thus be established prior to chondrule formation and chondrite accretion (Grossman, 1996). Disk models that address all the relevant physical processes are in their infancy, but recent theories have suggested ubiquitous, narrow regions in self-gravitating dusty 'clumps' well out in the disk that could melt aggregated dust (Lebreuilly et al., 2023).

Here, we provide new observations that further support the single formation reservoir relationship of matrix and chondrules in chondrite groups. Recent isotopic studies have established that the non-carbonaceous (NC, enstatite and ordinary chondrites) and CC groups differ in subtle ways (Trinquier et al., 2009; Warren, 2011). The consensus holds that the NC meteorites accreted inside and the CC outside of Jupiter's orbit (e.g., Kruijer et al., 2020). Here, we provide the first large-scale petrological and chemical comparison between NC (Semarkona, LL3.00) and CC (Renazzo, CR2) chondrites. In order to address the major unanswered question "Why are CAIs and chondrules sorted into different size and type populations within different chondrite types" (MacPherson et al., 2005, p.242), it is necessary to know with some accuracy those type abundances and sizes. We determine these abundances and sizes and extend the analysis to include the bulk chemical compositions of large numbers of individual objects.

Renazzo is one of only two CR chondrite falls, recognized by Wood (1963) to be an important record of early solar system processes. Weisberg et al. (1993, 1995) established the Renazzo CC type (CR) based on textural criteria (cf., McSween, 1977; King and King, 1978) in addition to bulk oxygen isotopes and AOA chemistry (Weisberg et al., 2004). The CR chondrites are distinguished macroscopically by abundant, often multilayered Type I (FeO-poor) chondrules with embedded or attached Fe-, Ni-alloy metal grains (Ebel et al., 2008). Although all CR2s experienced significant hydrous alteration, they experienced minimal thermal alteration (Krot et al., 2002; Abreu et al., 2020). McSween (1977) noted the strongly bimodal distribution of olivine compositions in Renazzo as an indicator of a lack of chemical equilibration. Here, we present new quantitative data on relatively large areas (total 1064 $mm^2$) of the Renazzo CR chondrite.

We also present new quantitative data on large areas (total 283 $mm^2$) of Semarkona (LL3.00, fall). Semarkona is the least equilibrated ordinary chondrite (Grossman and Brearley, 2005), with only minor evidence of aqueous and thermal metamorphism (Huss et al., 1981; Alexander et al., 1989; Alpert et al., 2021). The LL chondrites are most tractable among the ordinary chondrites (OC) due to the existence of low petrologic grade falls and their larger chondrules relative to L and H chondrites. The ordinary chondrites have been observed to record more enhanced recycling processes and interactions between chondrule melt and SiO- and Mg-rich gas than the carbonaceous chondrites (Piralla et al., 2021; Barosch et al., 2019).

## 2. PREVIOUS WORK

Point counting with an optical microscope has been the primary source of measurements in existing review compilations of chondrule, refractory inclusion and matrix modal abundances and object sizes (e.g., Grossman, 1988; Weisberg et al., 2006; Rubin, 2010). Methods and criteria for measuring matrix and inclusion abundances and sizes differ markedly among previous reports, particularly definitions of matrix and classification of isolated mineral grains. Here, we apply a unified classification scheme and consistent image analysis techniques at higher spatial resolutions than traditional optical methods across the least equilibrated representatives of the CR and LL chondrite groups. We also compare our results to data for CO and CV chondrites (Ebel et al., 2016), K chondrites (Barosch et al., 2020b), and CM2 chondrites (Fendrich and Ebel, 2021) also measured by image analysis. Preliminary iterations of the current work were published as abstracts (Bayron et al., 2014; Lobo et al., 2014).

## CR carbonaceous chondrites

Modal abundances of inclusions and matrix of Renazzo reported in the literature are initially due to McSween (1977, by optical microscopy point counting, following Dodd, 1976; cf. Chayes 1956). King and King (1978) reported 66 chondrules and 47.84% matrix (<0.1 mm particles) in 90 mm$^2$ of Renazzo (their CV2) and carried out a detailed chondrule size analysis. Bischoff et al. (1993) found about 60 vol% chondrules and chondrule fragments in Acfer 139 (CR2). Weisberg et al. (1993) point counted CRs by identifying chondrules, matrix, etc. on a grid placed over an optical photomosaic. Weisberg et al. (1993) also performed mineral modal analysis (olivine and pyroxenes) by automated 2-second electron probe microanalysis (EPMA) on a grid by stage motion with mineral identity determined in code, in an early application of automation to the problem of modal abundance based on Prinz et al. (1980; c.f., Delaney et al., 1983). Noguchi (1995) obtained modal abundances of components in PCA 91082 (CR2) by point counting in transmitted light (*n*=2900 gridded points), obtaining (matrix + dark inclusions)/(chondrules + fragments) = 0.49, and 0.7% refractory inclusions. He included 30-50µm fragments with matrix. McSween (1977, his Table 2) found 0.3 area% CAIs in Renazzo from 1562 points optically counted. Hezel et al. (2008) found 0.09 area% CAIs in 112 mm$^2$ of Renazzo with a log-normal size distribution and used statistical methods to deduce that CAIs must be about 0.12 area% in CR chondrites. Krot et al. (2002) state that 3-5 anorthite-rich chondrules (ARC) occur per thin section in CRs. Kallemeyn et al. (1994) reported "a few vol%" of 1-4 mm severely aqueously altered clasts in the Renazzo, MAC87320 and Al Rais (CR2) breccias. With the exception of Al Rais, these clasts generally correspond to the magnetite- and sulfide-rich xenolithic fragment described by McSween (1977), and the "dark inclusions" described by Weisberg et al. (1993).

More recently, modal abundances in CR chondrites were reported by Patzer et al., who measured clast modal abundances in CO (2021), CR (2022) and CM (2023) chondrites by point-counting on BSE images using grids of points (*n* ~1000) at 150 (CO, CM) or 300µm (CR) spacing. Each grid point was also analyzed quantitatively (3 µm spot, 15kV, 5nA, 1min dwell) and corrected against standards using software. Such a grid is superposed over a map of Renazzo in supplemental Figure S1 for comparison of methods. Schrader et al. (2011) used commercial software on BSE maps to obtain chondrule, matrix, and CAI abundances. Their method seems closest to that employed here in that it is a form of pixel-by-pixel image analysis by thresholding of BSE maps.

Evidence of petrofabrics in Renazzo and PCA 91082 (CR2) were reported by Kallemeyn et al. (1994) as a lineation "readily apparent in most thin sections" and measured in section USNM 6226-1. Charles et al. (2018) reported post-accretionary shape orientation from CT data on NWA 801 (CR2), with an inference of pre-accretionary non-sphericity. However, Noguchi (1995) reported that for PCA 91082 "the PTS investigated in this study does not show a lineation", nor has a petrofabric as reported in Renazzo by others. Distinct alignment of clasts was not observed in the present study; however, evidence of parent body shear forces has been observed in other sections (e.g., 588-t1-sA1-ps1A, Fig. 2). Such effects are not expected to have dramatically affected the sizes, relative abundances, or major element chemical compositions of inclusions and matrix in Renazzo.

Sizes of chondrules in CR chondrites are large relative to other carbonaceous chondrites (Friedrich et al., 2015). Kallemeyn et al. (1994) found a mean diameter of 690 (max 840, min 380) microns (n = 50) in Renazzo chondrules. Bischoff et al. (1993) found 1000 (±600) in 188 CR chondrules. Rubin (2000) used 700 µm as a best mean diameter. A complicating factor is the presence of fine-grained rims on chondrules (Metzler and Bischoff, 1996; Cosarinsky et al. 2008; Huss et al. 2005; Rubin 2010).

### LL ordinary chondrites

Modal abundances in ordinary chondrites were measured most carefully by Huss et al. (1981) who concluded they contain 60%-80% chondrules, 10%-15% metals/opaque minerals, and the remainder fine-grained matrix. These values occur in many reviews of meteorite properties (e.g., Weisberg et al., 2006). Most importantly, Huss et al. (1981) clearly defined identification criteria for each component including matrix. Gooding and Keil (1981) provided clast abundance data for Semarkona. Grossman and Brearley (2005) used EPMA elemental maps to determine matrix/inclusion/metal/sulfide ratios by X-ray map image analysis of ordinary chondrites including Semarkona.

## 3. METHODS

### Overview

Our goal was to measure 1) the relative proportions of objects (clasts) of different bulk chemical composition and texture (e.g., CAIs, AOAs, chondrules, metal-rich opaque assemblages), 2) clast/matrix ratios, and 3) the distributions of major elements among each of these components. We achieved this by using X-ray intensity measurements on one-micron spots evenly spaced across large areas. These X-ray element maps were then used to digitally outline each object (clast) in each area. We then used our ability to query each pixel in each clast to extract X-ray intensity statistics for the major elements in each clast and in the bulk matrix and converted the overall intensity values for each clast and total matrix to weight %. These results allow assessment of the chemical variability among clast types and matrix. We discuss these findings in the context of contributions of clasts and matrix to the bulk chondritic elemental abundances. In analyzing large numbers of clasts across large surface areas, we have endeavored to measure the overall characteristics of the forest of clasts, without focusing on potentially fascinating individual inclusions.

## Samples

Samples were chosen from the American Museum of Natural History (AMNH) meteorite collection (Table 1). Both are observed falls. Complementary 3D data exist for three slabs that were scanned using computed tomography (CT) prior to cutting for sections. We do not use any CT data in the present work. CT and EPMA data are curated and available for study as part of the AMNH meteorite collection.

A polished ~6mm thick slab of Renazzo (4905A) was chosen for study and CT scanned. Both sides (ps1A, ps1B) were mapped using EPMA. A smaller slab (588-t1) was CT scanned at 17.1 µm/voxel edge (Ebel and Rivers, 2007) and a surface of a smaller section (588-t1-sA1) was mapped (588-t1-sA1-ps1A; cf., Ebel et al., 2008). Results for the three Renazzo sections, all diamond-polished, were combined for component abundance analysis but the 588-t1 sample was not used for clast-by-clast elemental analyses as it was not mapped for the same suite of elements as the other two sections.

**Table 1:** Samples and imaging conditions. Resolution (res) in micrometers/pixel, dwell time (ms), number of $512^2$ frames (n), corresponding area (A, $mm^2$), WDS and EDS elements, and the total number of non-masked pixels in each map are stated.

| meteorite | AMNH sample | res | ms | kV | nA | n | A($mm^2$) | pass 1: WDS | pass 2: EDS | total pixels |
|---|---|---|---|---|---|---|---|---|---|---|
| Renazzo | 4905A-t1-ps1A | 6 | 20 | 15 | 20 | 60 | 538 | Mg,Al,Ca,Ti,Ni | Mg,Si,S,Fe | 14943350 |
| Renazzo | 4905A-t1-ps1B | 5 | 25 | 15 | 20 | 72 | 454 | Mg,Al,Ca,Ti,Ni | Mg,Si,S,Fe | 18156210 |
| Renazzo | 588-t1-sA1-ps1A | 4 | 20 | 20 | 20 | 21 | 72 | Si,Fe,Ca,Mg,Al | none | 4525787 |
| | | | | | | | | | | |
| Semarkona | 4128-5 | 10 | 15 | 15 | 20 | 8 | 157 | Mg,Ni,Ti,Al,Ca | Na,Si,S,K,Fe | 1568407 |
| Semarkona | 4128-t1-ps1B | 8 | 5 | 15 | 40 | 10 | 127 | Si,Fe,Ca,Mg,Al | S,Mn,Ni | 1978380 |

A doubly-polished, commercially prepared thin section of Semarkona, AMNH 4128-5 was chosen for analysis. A ~6 x 8 x 22 mm (2.32 g) irregular slab of Semarkona (AMNH 4128-t1) was CT scanned, then cut nearly parallel to a cut face (ps1A) with a boron carbide slurry on a 50 µm tungsten wire saw to expose face ps1B in a ~200 µm thick slab, and a new face of the remaining slab, ps2A. Diamond-polished sample 4128-t1-ps1B was analyzed here. Results for the thin and thick sections (Table 1) were combined for presentation.

## Mapping

Intensity of Kα X-ray emission for selected elements was measured in raster maps using the Cameca SX100 electron probe microanalyzer (EPMA) at the AMNH, with five wavelength dispersive spectrometers (WDS) and equipped with a Bruker AXS Quantax 4010 energy dispersive spectrometer (EDS). Map conditions are listed in **Table 1**. Back-scattered electron (BSE) images were collected for all samples. All maps were made using a 1 µm spot beam in stage mapping mode keeping the beam nominally perpendicular to the surface. Mosaics of many adjacent, usually 512x512 pixel frames were collected on each sample surface by rastering stage motion in each frame with the beam focused at the frame center. Map areas were set to overlap by one or two pixels on each edge. Background counts were not routinely collected so as to maximize pixel resolution and mapped area within allotted instrument time (Crapster-Pregont and Ebel, 2020; Fendrich and Ebel, 2021). Data for minor elements Ti, K, Na, and Mn did not yield sufficient counts to be quantitatively robust. Spatial resolution and dwell time of each mosaic were selected to optimize resolution and sample coverage within the EPMA time available (e.g., mapping 4905A-t1-ps1B took 182 hours). Both 8-bit and 32-bit TIF format files were output, the latter used for all quantitative analysis. Stitched together 8-bit and 32-bit element map mosaics and color-balanced, three-element 24-bit red-green-blue (RGB) composites (e.g., Mg-Ca-Fe) were produced using custom software. Versions of element maps in 8-bit format are presented as supplemental Figures S2.

*Matrix Analysis*

Previous investigators have analyzed matrix for major elements using large, defocused EPMA beam spots. McSween and Richardson (1977) used a 100µm beam (20kV, 40nA, 20s) to analyze matrix in 32 CCs. They noted that "weight per cent Mg/Si in matrices is constant (0.82 ± 0.05) but less than ratios derived from bulk analyses". They compared Allende analyses to wet chemical analysis of a <100 µm sieved fraction by Clarke et al. (1970), reporting that "agreement between the two methods of analysis is generally very good". Grossman and Brearley (2005) used rastered beam motion (e.g., 30x30µm) with a 1µm beam. Zolensky et al. (1993) used a focused beam and noted that their results compared well with those reported by McSween and Richardson (1977). Huss et al. (1981) tested EPMA methods on artificial matrix and concluded that “none of the errors is large enough to affect the conclusions reached” in their work. However, Warren (1997) found that broad beam EPMA analysis generally yields concentrations of Mg and Fe too low by ~1.22x and Al concentration too high by the same factor.

Hezel and Palme (2010) found complementary Mg/Si in CV, CR, CO and CM chondrites with EPMA (15kV, 20nA, 1µm olivine + pyroxene, 5µm feldspar, 30µm matrix, ZAF applied), fine-grained rims avoided. They noted that "absolute concentrations from broad beam matrix analyses are unreliable as matrix can be porous and contains sulfide and metal" which are calculated as FeO, but that "the Mg/Si ratios are unaffected by these analytical limitations". Their bulk chondrule compositions were calculated by modal reconstruction from spot analyses and modes derived from element maps with an "absolute error" correction applied to 2D data based on Hezel and Kießwetter (2010).

Zanda et al. (2018) pointed out that "caution needs to be exercised when data obtained from different analytical methods are used as a basis for understanding chondritic compositions", based on a comparison of broad beam EPMA (their work on Orgueil, 10µm spot, 15kV, 10nA) with wet chemical results. They found *low* apparent

Mg/Si ratios by EPMA relative to wet chemistry. Comparison with CI chondrites is an extreme case, as these are the most volatile-rich and the most porous chondrites. If anything, EPMA Mg/Si mass ratios for matrix, while below bulk values, are too *high* relative to what would result from wet chemical analysis, as is demonstrated through comparisons of less porous, volatile depleted CV chondrite data acquired using different methods in Table 2. The question of bias in matrix EPMA analysis among different chondrite types with different matrix properties remains open.

**Table 2:** Comparison of Si, Mg and Al abundances in bulk and matrix of Allende (CV3) for three techniques. Broad beam EPMA of matrix yields a higher Mg/Si ratio than the only known wet chemical analysis (Clarke et al., 1970, analysis by Jarosewich) but a lower Al/Si ratio. Data are from Clarke et al. (1970, C+ 70), Jarosewich et al. (1987, J+ 87), Stracke et al. (2012, S+ 12), McSween and Richardson (1977, MR 77), and Kimura and Ikeda (1998, KI 98).

| | **reported oxide wt.% composition** | | | | | | | **recalculated to atom wt% from oxide wt.%** | | | | | |
|---|---|---|---|---|---|---|---|---|---|---|---|---|---|
| **method:** | **wet chemistry** | | | **XRF** | **broad beam EPMA** | | | **wet chemistry** | | | **XRF** | **broad beam EPMA** | |
| **source:** | C+ 70 bulk | J+ 87 bulk | **C+ 70 matrix** | S+ 12 bulk | MR 77 matrix | KI 98 matrix | | C+ 70 bulk | J+ 87 bulk | **C+ 70 matrix** | S+ 12 bulk | MR 77 matrix | KI 98 matrix |
| SiO2 | 34.23 | 34.27 | **33.11** | 34.24 | 28.00 | 27.90 | Si | 16.00 | 16.02 | **15.48** | 16.00 | 13.09 | 13.04 |
| MgO | 24.62 | 24.51 | **21.42** | 24.77 | 20.20 | 19.00 | Mg | 14.85 | 14.78 | **12.92** | 14.94 | 12.18 | 11.46 |
| Al2O3 | 3.27 | 3.35 | **3.07** | 3.00 | 2.30 | 2.10 | Al | 1.73 | 1.77 | **1.62** | 1.59 | 1.22 | 1.11 |
| | | | | | | **Element ratios** | | | | | | | |
| | | | | | | Mg/Si mass ratio | | 0.928 | 0.923 | **0.835** | 0.933 | 0.931 | 0.879 |
| | | | | | | Al/Si mass ratio | | 0.108 | 0.111 | **0.105** | 0.099 | 0.093 | 0.085 |

In our approach, if we have correctly segmented maps to isolate inclusions from matrix, then our matrix analysis technique is identical to rastering over a very large area. As pointed out by Grossman and Brearley (2005), matrix grain size is small relative to the excitation volume of the electron beam, reducing the need for density corrections. Here, fine-grained chondrule rim material is not differentiated from matrix but, in Renazzo, coarser rim material compositionally and texturally like the host chondrule was outlined as part of that chondrule. We provide maps and outlines illustrating this as supplemental Figures S1 to S3).

*Map calibration*

X-ray intensity EPMA maps were collected on a total area on Renazzo of 992 $mm^2$. Conversion of element counts per pixel to weight percent (wt%) was done by assuming that for each element, the bulk (e.g., 32678953 pixels on Renazzo) mean count/pixel corresponds to the chemically determined bulk wt% composition of Renazzo (Mason and Wiik, 1962). We have demonstrated elsewhere that calibration curves correlating average counts/pixel to known wt% on known standards are linear (Crapster-Pregont and Ebel, 2020).

X-ray intensity EPMA maps were collected on a total area of 283 $mm^2$ of Semarkona. Conversion of average element counts per pixel to wt% was done by assuming that the bulk count/pixel on all pixels corresponds to the chemically determined bulk wt% composition (Jarosewich, 1966) of Semarkona for each section. For each outlined object, average counts per pixel were converted to bulk wt% for that object using the same conversion factors as for the bulk composition, as for Renazzo.

## Image Analysis

The workflow for image analysis closely follows that of Ebel et al. (2016), which presents a much more complete description (cf., Crapster-Pregont and Ebel, 2020; Fendrich and Ebel, 2021). ImageJ (ImageJ, 2024) and custom software written in IDL (IDL, 2024) were used to determine clast area abundances, apparent clast sizes, and the major element composition of each individual clast and of all the matrix pixels as an aggregate.

Segmentation, the identification of the boundary of each non-matrix object in each map, was performed using drawing tools in the Adobe Illustrator™ (AI) software. All the element maps and several red-green-blue (RGB) composites were imported as registered (exactly overlayed) layers into AI (supplemental Figures S2). The parts of images that did not contain sample were masked off. Outlines were drawn by hand in AI as vector graphic objects (curves thinner than pixels in the element maps) around each of the clasts present in each surface. Identifiable clast types were separated into AI layers with specific grayscale fill values (Ebel et al., 2016, their Fig. 1E). For Semarkona (LL3.0) the silicate clasts were outlined and measured separately from the opaque materials due to their close proximity in order to avoid them being joined in ImageJ, which imports bitmap versions of the AI output files, with vector objects converted to pixelated shapes.

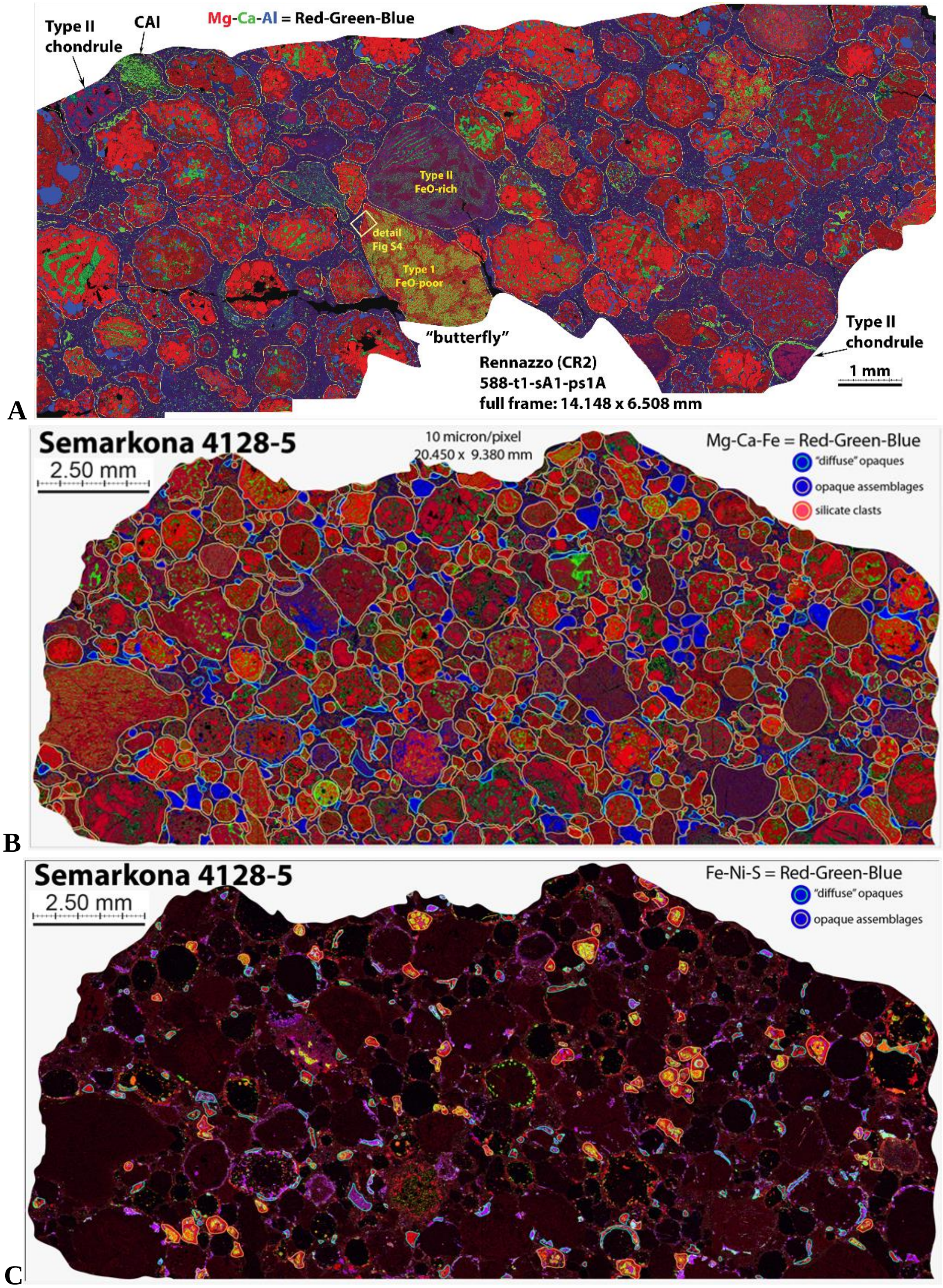

**Figure 2: A:** Renazzo sample 588-t1-sA1-ps1A in Mg-Ca-Fe red-green-blue mosaic. Clasts, primarily chondrules, are outlined in white. Two remarkable barred olivine chondrules in center, collectively called the "butterfly", are FeO-rich Type II and FeO-poor Type I and exhibit dramatic evidence of in situ shear forces (supplemental Figure S4). **B:** Semarkona sample 4128-5 in Mg-Ca-Fe=RGB mosaic. Three types of material are outlined. **C**: 4128-5 in Fe-Ni-S=RGB showing "OAs" and "diffuse" opaques.

Thresholded grayscale maps in ImageJ allowed numbering each discrete identified object to produce a table with a bounding rectangle for each enumerated object. Within each rectangle, the IDL code “grew” each object pixel-by-pixel to its limit, excluding the surrounding matrix. This yielded a map with each object’s pixels having a discrete R-G-B color, also identified in a table. The combination of maps and tables allowed software to address the pixels in each clast individually across all the element maps. Supplemental Figures S1 to S3 include the RGB maps, and supplemental Tables S1 identify all objects present in image analysis maps. Algorithms and code are further detailed in the supplemental online material of Ebel et al. (2016).

Chondrule clasts were identified as barred olivine (BO), radial pyroxene (RP), or porphyritic by inspection. CAIs and AOAs in Renazzo, and opaque assemblages (OA) in Semarkona were also identified by inspection. Two-dimensional data are insufficient to characterize porphyritic chondrules as olivine- or pyroxene-dominated so all chondrules not BO or RP are categorized as “porphyritic” (Barosch et al., 2019, 2020a). We did not distinguish between Type I (FeO-poor) and Type II (FeO-rich) chondrule types, although statistics on the two populations could easily be extracted from our dataset. In the LL chondrite, more “diffuse”, sulfide-rich Fe-rich material was segmented but combined with matrix. Discrete, coherent, consolidated “metallic chondrules” (Gooding and Keil, 1981), corresponding to the opaque assemblages (OAs) of Alpert et al. (2021), were separately outlined in LL chondrites.

Segmented (outlined) maps allowed computation of the relative proportions of each object type, as well as the total area measured, and the fraction of matrix. Each pixel in each object was then analyzed for the relative abundances of elements measured in the X-ray maps. Pixels with a low sum of major element X-ray intensities were labeled as holes, and not considered further (supplemental Tables S2). The total intensity of emission for each element over the area of each object, minus holes, was computed separately for each inclusion and in aggregate (minus holes and cracks) for matrix. From this data, the average wt% of major elements in each object, and bulk matrix, could be computed.

## 4. RESULTS

### Inclusion and matrix abundance

Modal abundances of components in Renazzo obtained in this study are nearly identical to those of McSween (1977), who counted metal-rich chondrules separately, as we did (**Table 3)**. Refractory inclusions (AOA and CAI) are sufficiently rare in CR chondrites that significant variation between sections is to be expected, making traditional point counting methods a poor choice for accurate characterization of their abundances. Only a single radial pyroxene (RP) chondrule was observed across all the area measured in Renazzo, in contrast to 15 in Semarkona. Rubin (2010) also noted fewer RP and cryptocrystalline chondrules in CCs relative to OCs. Renazzo also has much fewer BO chondrules per area, relative to Semarkona.

Our modal abundances of matrix and inclusions in Renazzo are consistent with previous work based on point counting (McSween, 1977; King and King, 1978; Weisberg et al., 1993; Schrader et al., 2011; Patzer et al., 2022). Here, our area coverage is significantly larger with much closer spacing of points (pixels). Our segmentation may slightly overcount matrix material due to the vector outline vs bitmap issue noted above. Our result for AOAs and CAIs, with AOAs ~2x more abundant than CAIs, is strikingly close to that of McSween (1977), not exceedingly different from the Poisson distribution of Hezel et al. (2008). Our comparison with earlier work re-establishes that the number of points counted, and their spacing, is critical to the accurate measurement of the abundance of less abundant components (e.g., CAIs, AOAs; c.f., supplemental Figure S1).

**Table 3:** Area percent of component types, compared to previous work on Renazzo (CR2). Only RP, BO, and porphyritic chondrule types are distinguished in the current work. The total of matrix, lithic and mineral fragments and isolated opaque phases is divided by the total of CAI, AOA and chondrules to compute the matrix/inclusion ratio.

| Renazzo results | M77 | KK78 | W+93 | S+11 | P+22 | this work |
|---|---|---|---|---|---|---|
| *matrix* | 31.1 | | 37.5 | 27.6 | 38.9 | 33.33 |
| *lithic fragments* | 1.9 | | 7.5 | | | 3.23 |
| *mineral fragments* | | | | | 0.4 | 0.32 |
| *opaque phases* | 5.5 | | | | 0.7 | 0.48 |
| CAI (or total RIs*) | 0.3 | | 1 | 0.2 | 0.1 | 0.43 |
| AOA | 1.3 | | | | 2.2 | 1.13 |
| RP chondrules | | | | | | 0.23 |
| BO chondrules | | | | | | 2.41 |
| porph/all chondrules | 25.6 | | 54 | 72.2 | 57.7 | 58.43 |
| opaque-rich chon. | 34.4 | | | | | |
| *sum of matrix* | 38.5 | 47.84 | 45.0 | 27.6 | 40 | 37.35 |
| sum of inclusions | 61.6 | 52.16 | 55.0 | 72.4 | 60 | 62.65 |
| *matrix* /inclusions | 0.625 | 0.917 | 0.818 | 0.381 | 0.667 | 0.59 |
| chondrule number | | 66 | | | | 1255 |
| CAI number | | | | | | 61 |
| AOA number | | | | | | 35 |
| points counted | 1562 | | | n.a. | 1054 | 37625347 |
| point spacing (μm) | ~275 | | | n.a. | 300 | 5.277 |
| area analyzed ($mm^2$) | | | 256.7 | 273.9 | 94.86 | 1064.3 |

**Notes on Table 3:**

*Sum of matrix* is the sum of the first four lines in italics. M77: McSween (1977) did not report the measured area nor retain records (pers comm.). KK78: King and King (1978) only reported the total matrix (<0.1mm) fraction. W+93: Weisberg et al. (1993) include lithic and mineral fragments and isolated opaques in matrix, and (*) combined CAI and AOA as refractory inclusions (RI), as did Schrader et al. 2011. S+11: Schrader et al. (2011), this is their result for Renazzo only (of 20 CRs reported). P+22: Patzer et al. (2022), this is the weighted average for 2 Renazzo areas, the total area is calculated.

**Table 4:** Area percentage of component types, compared to previous work on Semarkona (LL3.00). Only RP, BO, and porphyritic chondrule types are only distinguished in the current work. The total of matrix, lithic and mineral fragments and isolated opaque phases is divided by the total of CAI, AOA and chondrules to compute the matrix/inclusion ratio. We found no CAIs or AOAs in Semarkona but small CAIs are known to exist (Bischoff and Keil, 1984; Huss et al., 2001).

| **Semarkona results** | H+81 | GK81 | GB05 | this work |
|---|---|---|---|---|
| *matrix* | 15.6 | 23* | 16.1 | 21.81 |
| *lithic fragments* | | | | 0.98 |
| *mineral fragments* | | | | 0.00 |
| *opaque phases* | | | 3.9 | 2.83 |
| CAI and AOA | | | | 0.00 |
| RP chondrules | | 7.7 | | 4.54 |
| BO chondrules | | 2.31 | | 4.45 |
| porph/all chondrules | | 65.45 | 76.9 | 61.21 |
| OA opaque assemblages | | 0 | | 4.18 |
| *sum of matrix* | | | | 25.62 |
| sum of inclusions | | | | 74.38 |
| *matrix* /inclusions | | | | 0.344 |
| chondrule number | | 213 | 167 | 781 |
| OA number | | | | 104 |
| points counted | | | | 3546787 |
| mean point spacing (μm) | | | | 9.12 |
| area analyzed ($mm^2$) | | | 60 | 298.216 |

**Notes on Table 4:**
*Sum of matrix* is the sum of the first three lines in italics. H+81: Huss et al. (1981) reported only matrix fraction. GK81: Gooding & Keil (1981) reported fractions of chondrule types, here summed (*) to 77% of their total. GB05: Grossman & Brearley (2005) reported only matrix abundance and opaques, and computed chondrules by difference.

**Object sizes**

Apparent size distributions of chondrules, opaque assemblages (OA), AOAs, and CAIs were determined from the outlined objects in maps for each chondrite surface. Sizes were calculated as the diameter of a circle of equivalent area to each object. We present these results in terms of diameter histograms rather than the phi, $\phi = -\log_2(\text{diameter, mm})$, scale common in sedimentary geology (Folk, 1980). Isolated mineral grains in matrix were not included in size determinations. We do not attempt to convert apparent, 2-dimensional (2D), size determinations to 3D distributions. Friedrich et al. (2022) determined true 3D sizes of Semarkona chondrules using computed CT methods and reviewed recent work that characterized the relationship between 2D and 3D chondrule sizes (e.g., Metzler, 2018).

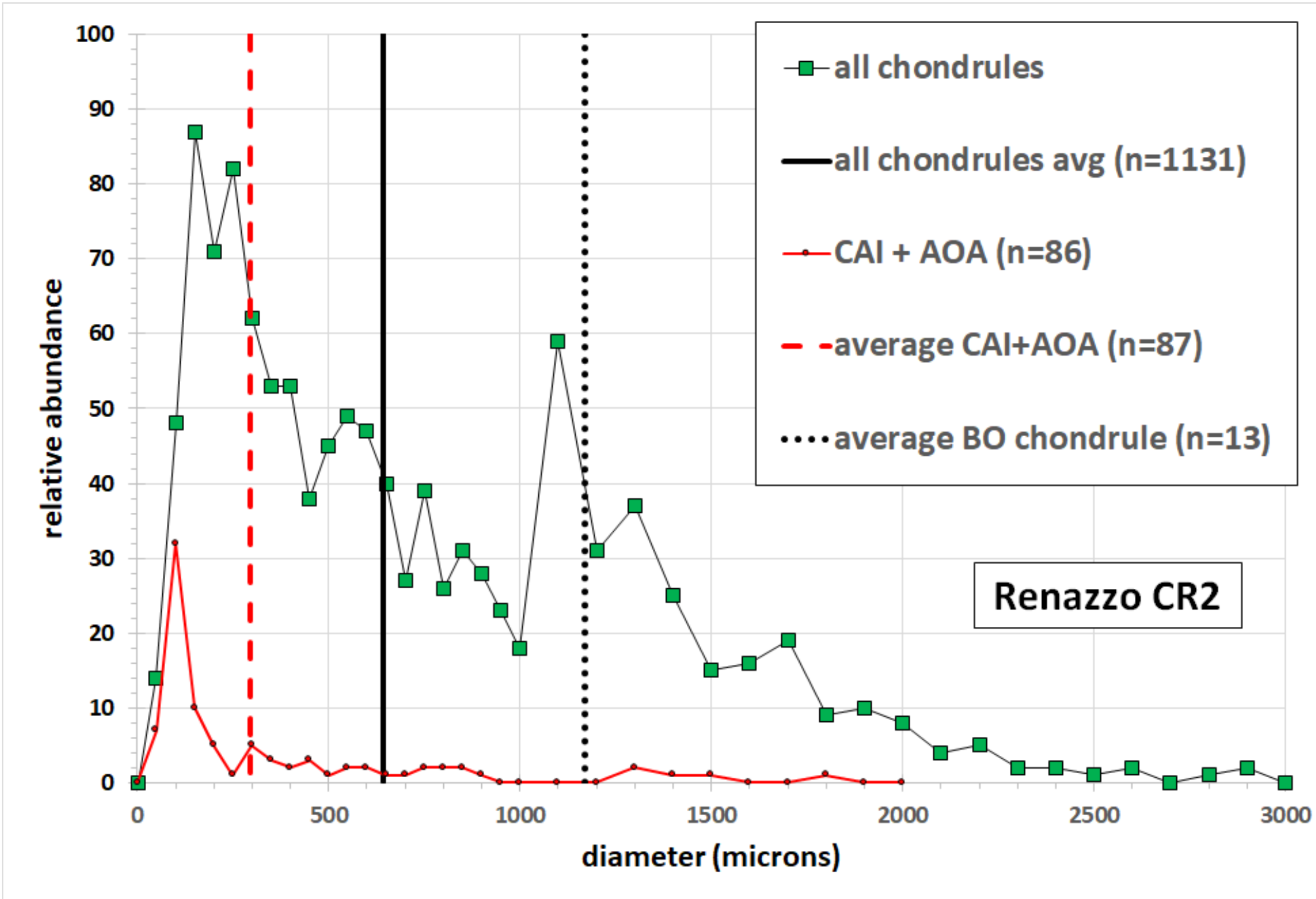

**Figure 3:** Size histogram of Renazzo chondrules and refractory inclusions, with average sizes for chondrules and refractory inclusions (CAI + AOA) and barred olivine (BO) chondrules. Apparent diameters are for circles equivalent to 2D area measurements.

In Renazzo, we find that the mean diameter of 1131 chondrules is 617 µm, 1σ=510, range 3412 to 21 µm. The AOA (n=29) and CAI (n=58) mean diameter is 294 µm, and the barred olivine (BO) chondrules (n=13) average 1168 µm (Fig. 3). Only one radial pyroxene chondrule was found in Renazzo. Weisberg et al. (1993) reported a mean size of 830 µm for 169 Renazzo chondrules, while Kallemeyn et al. (1994) found a mean size of 690 µm (+840/−380) in 50 Renazzo chondrules using optical methods.

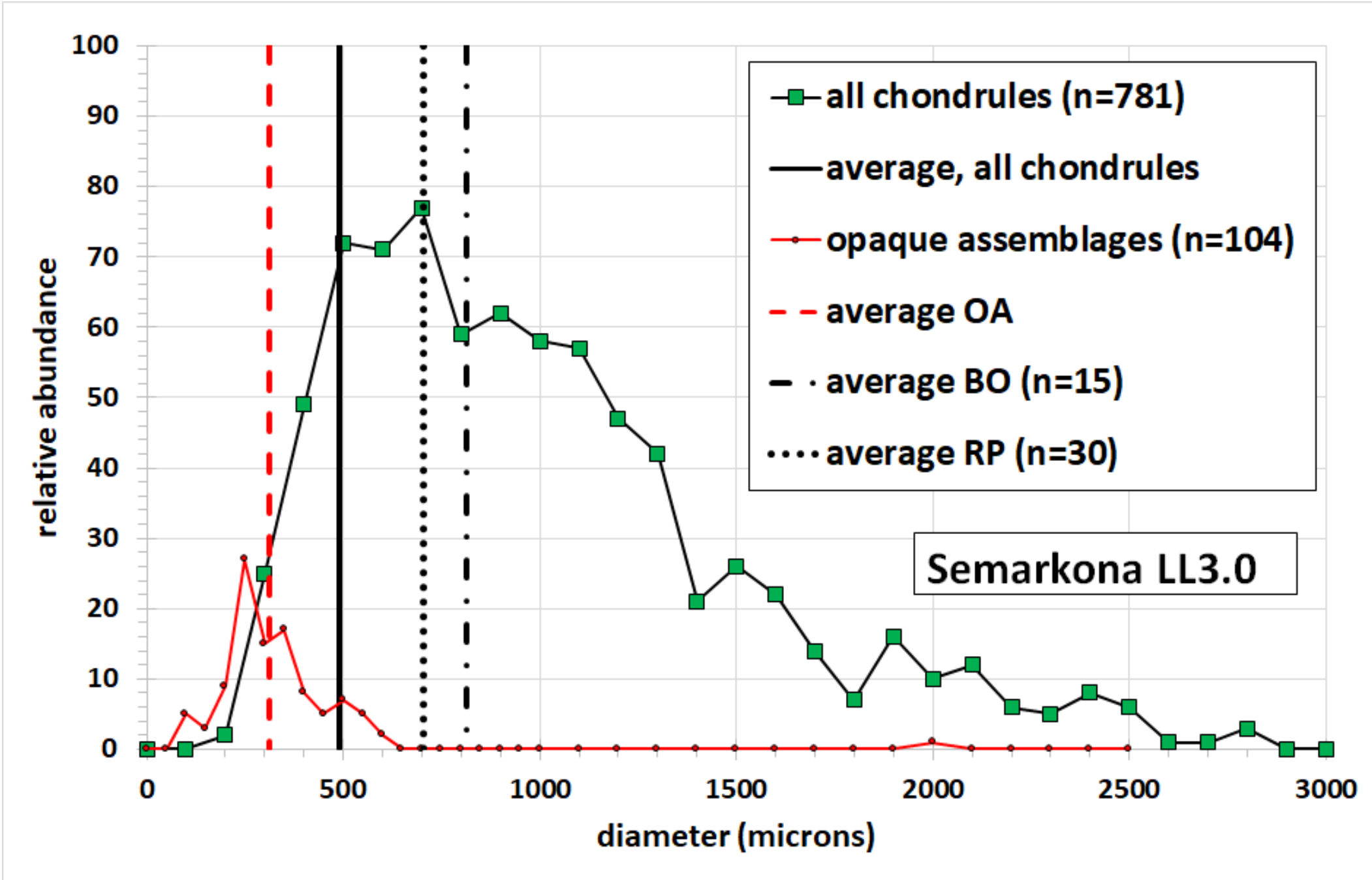

**Figure 4:** Size histogram of Semarkona chondrules and opaque assemblages (OA) with average sizes for chondrules and OAs and barred olivine (BO) and radial pyroxene (RP) chondrules. Apparent diameters are for circles equivalent to 2D area measurements.

In Semarkona, we find that the mean diameter of 781 chondrules is 490 μm, 1σ=295, range 76.5 to 2493 μm (Fig. 4). The radial pyroxene (RP) and barred olivine (BO) chondrules are significantly larger than the average chondrule, averaging 704 μm (n=30) and 814 μm (n=15), respectively. Gooding (1983) determined sizes of Semarkona chondrules, finding geometric mean sizes 1390 microns (+890/-540 standard deviations) for 15 chondrules, while Huang et al. (1996) found a mean size of 753 μm (338 1σ) for 190 chondrules, both in polished thin sections using optical microscope reticules (Friedrich et al., 2015). Nelson and Rubin (2002) found a mean diameter of 570 μm for 719 intact chondrules in LL chondrites, with “droplet chondrules” (BO, RP, cryptocrystalline) also larger. It appears that as the number of objects counted increases, the mean diameter decreases, likely due to more systematic measurement of small chondrules.

## Composition of components

### *Complementary Mg-Si relationships*

The methods employed here allow direct addition of the x-ray intensities of each element in each pixel in each inclusion (clast), and in the aggregate of matrix pixels. The sum over the total number of pixels in each inclusion, or all matrix pixels, or over the entire section may be compared. It is also straightforward to combine categories or separate them during analysis. Complete data sets and our analysis are available as supplemental data. Each individual chondrule or inclusion is represented by its bulk chemistry and weighted by its size, shown in weighted density distributions in Figures 5 through 8. The points representing groups of objects (e.g., “barred olivine”) are averages weighted by size for all those objects. For example, in Semarkona t1-ps1B, one large RP chondrule with FeO-rich olivine pulls the RP weighted average to higher wt% Si. The points for “all chondrules” overlap the weighted average for porphyritic chondrules in all cases.

The chondrules and matrix in both CR (Fig. 5) and LL (Fig. 6) chondrites illustrate distinct, opposing differences in their Mg/Si mass ratios from that of the bulk meteorite. For Renazzo (CR2), the bulk Mg/Si ratio by EPMA is very close to the Mg/Si mass ratio of bulk CI chondrites and that observed in the solar photosphere (Lodders, 2020). This relationship is demonstrative of the term “complementarity” (Palme et al., 2014). For Semarkona (LL3.0) the bulk Mg/Si ratio is ~90% of CI (Fig. 1), inconsistent with the definition of “complementarity” proposed by Hezel et al. (2018), which calls for a bulk composition closer to solar values, as will be discussed later. In these figures, the CI wt% values are lower due to their mass fraction dilution by more volatile elements, which are relatively depleted in CR and LL chondrites.

In the following Mg-Si plots, we have combined matrix with the rare lithic clasts commonly called dark inclusions (DI), and with isolated olivine grains (IOL) and small opaque grains in the matrix (FeNi). In Renazzo, Fig. 5, any combination of matrix, DI, FeNi, and IOL, or just the matrix, all plot within the same larger point (purple square) due to the volumetric dominance of matrix. Similarly, for Renazzo (Fig. 5), combining IOL with chondrules has a negligible effect on the mean chondrule composition. In both cases, consistent with the consensus that IOL derive from chondrules (e.g., Jacquet et al., 2021), combining IOL with chondrules would accentuate the chondrule-matrix difference. Combining AOAs and CAIs with the chondrules yields a point, “all clasts”, in essentially the same place as the weighted mean chondrule.

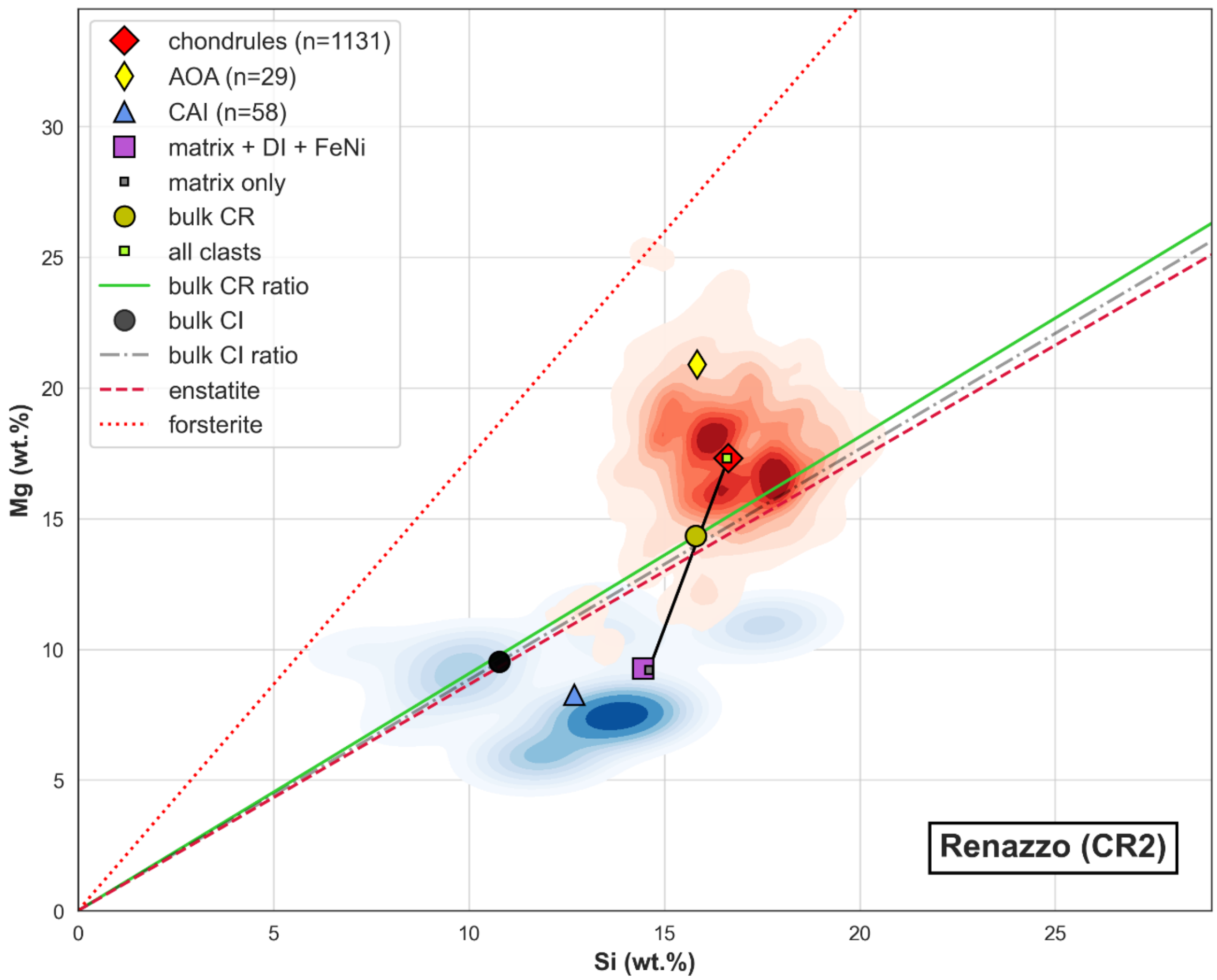

**Figure 5**: Wt% Mg and Si of inclusions and matrix in Renazzo, AMNH 4905A ps1A and ps1B (922 $mm^2$). Weighted density distributions represent the spread of individual chondrules (red) and CAIs (blue), with points showing weighted averages for each inclusion type, matrix, and bulk (all pixels). Control lines show Mg-olivine (forsterite) and Mg-pyroxene (enstatite) ratios.

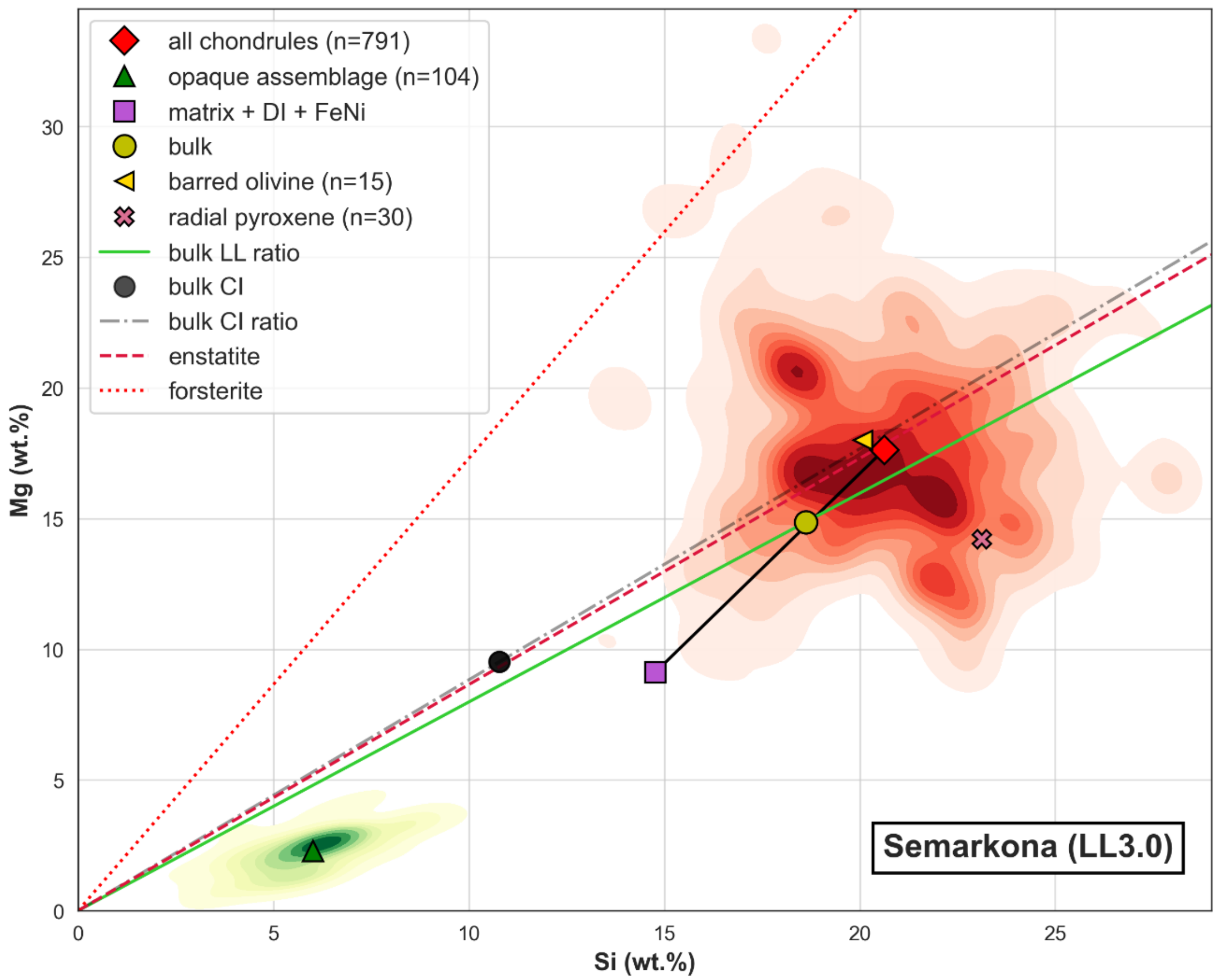

**Figure 6**: Wt% Mg and Si of inclusions and matrix in Semarkona. Weighted density distributions represent the spread of individual chondrules (red) and opaque assemblages (green), with points showing weighted averages for each inclusion type, matrix, and bulk (all pixels). The weighted mean of all 791 chondrules is the red diamond. Control lines show Mg-olivine (forsterite) and Mg-pyroxene (enstatite) ratios.

In Semarkona (Fig. 6), there is sufficient opaque, Fe-rich material in the matrix, separate from discrete, "metallic chondrule"-like OAs, that matrix is pulled slightly toward the origin in the Mg-Si plot. Any combination of matrix, DI, FeNi, and IOL, or just the matrix, all plot within the same point (purple square) due to the volumetric dominance of matrix among these components.

*Fe-Si relationships*

Analyses of these sections reveal similar relationships between Fe and Si in matrix and clasts in both meteorites. Because of the generally low FeO content of chondrule silicates, Fe behaves differently than Mg (Palme, 1992; Palme et al., 2014). Wood (1963) also remarked on this relationship between Fe/Si in chondrules and matrix in many chondrites.

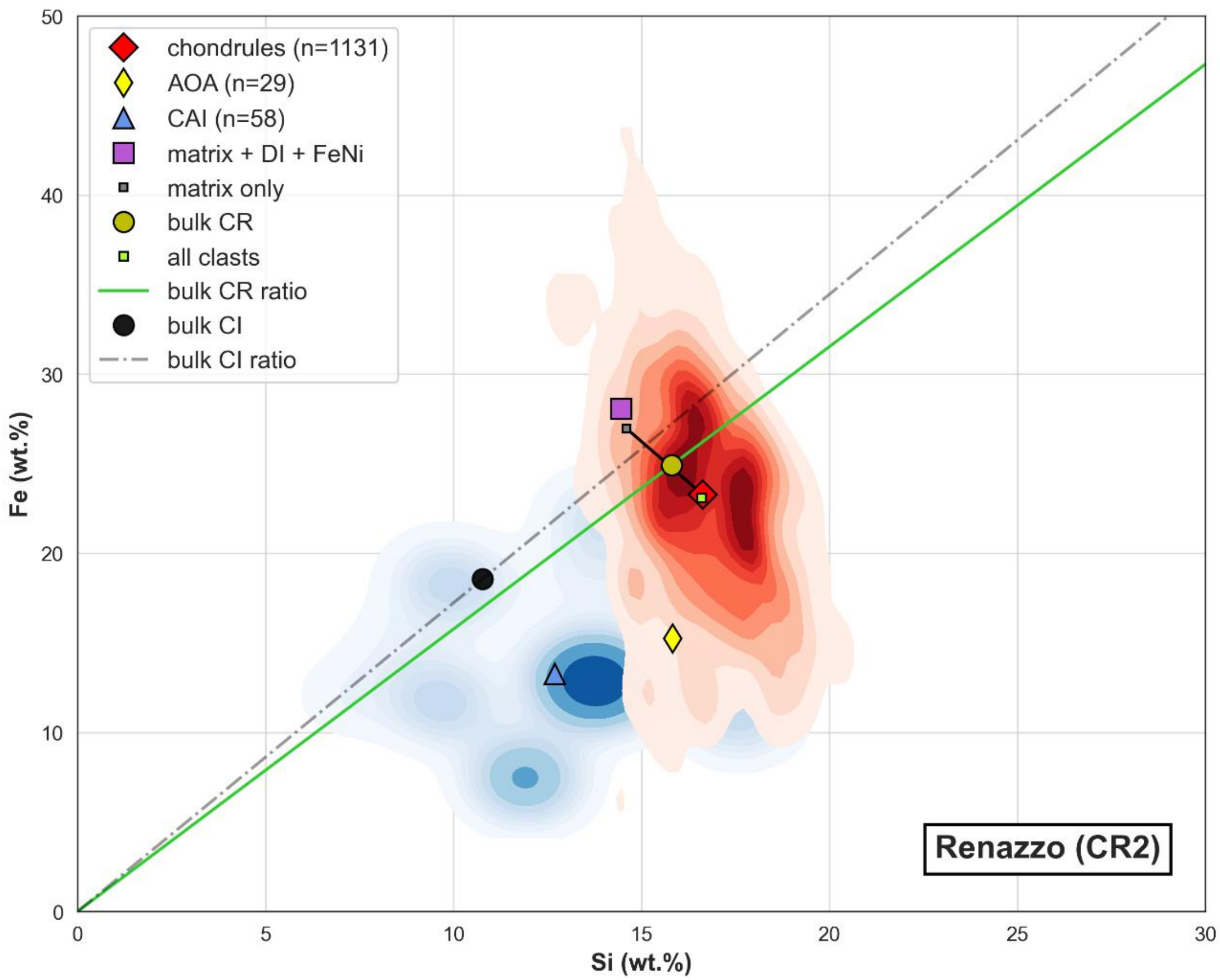

**Figure 7**: Wt% Fe and Si of inclusions and matrix in Renazzo. Weighted density distributions represent the spread of individual chondrules (red) and CAIs (blue), with points showing weighted averages for each inclusion type, matrix, and bulk (all pixels) The weighted mean of all 1131 chondrules is the red diamond.

We find that the Fe distribution among chondrite components in Renazzo is "complementary" as well, as did Kong and Palme (1999). This is evidenced by the matrix and chondrule averages plotting opposite one another across the bulk CR value in Figure 7 – meaning the matrix is enriched in Fe by the amount that the total inclusions are depleted. Even though much of the metal in CR chondrites is inside chondrules (e.g., Ebel et al., 2008), the matrix contains more total Fe than chondrules (Fig. 7). In all cases, the bulk falls in the triangle between other major components. In Renazzo, the CAIs and AOAs negligibly affect the bulk, which is volumetrically dominated by matrix and chondrules.

Expulsion of metal from CR chondrules has been hypothesized (Connolly et al., 2001; Jacquet et al., 2013). Combination of apparently isolated metal grains in Renazzo with the chondrules, rather than matrix, would bring the two points in Figure 7 closer together. This is illustrated in supplemental Figure S6a for Renazzo, with a similar grouping for Semarkona (supplemental Figure S6b).

*Bearing on opaque assemblages (OAs) in LL chondrites*

Opaque assemblages (OA) and Fe-rich matrix material strongly affect the bulk Fe-Si relationships in Semarkona (Fig. 8). The OAs are a significant carrier of Fe in the LL chondrites. The origin of OAs has been debated, with their common rimming by magnetite (Alpert et al., 2021) being considered a parent body phenomenon (Krot et al., 2022). Some have argued that metal-rich objects in matrix were "expelled" by chondrules as immiscible Fe-rich melts in a silicate melt, either by rapid chondrule rotation (Tsuchiyama et al., 2000) or by surface tension effects (Grossman & Wasson, 1985). We have seen no evidence in CT analysis of equatorial metal on chondrule rims that would be expected from rotation (Ebel et al., 2008). The large relative sizes of OAs relative to chondrules in Semarkona (Figs. 2, 4; Alpert et al., 2021) seems contrary to the expulsion hypothesis. It seems that a nebula origin of OAs, or at least their cores, in LL chondrites cannot be dismissed. Alpert et al. (2021) found striking differences between discrete OAs in matrix and the Fe-rich blebs found inside chondrules, supporting this concept.

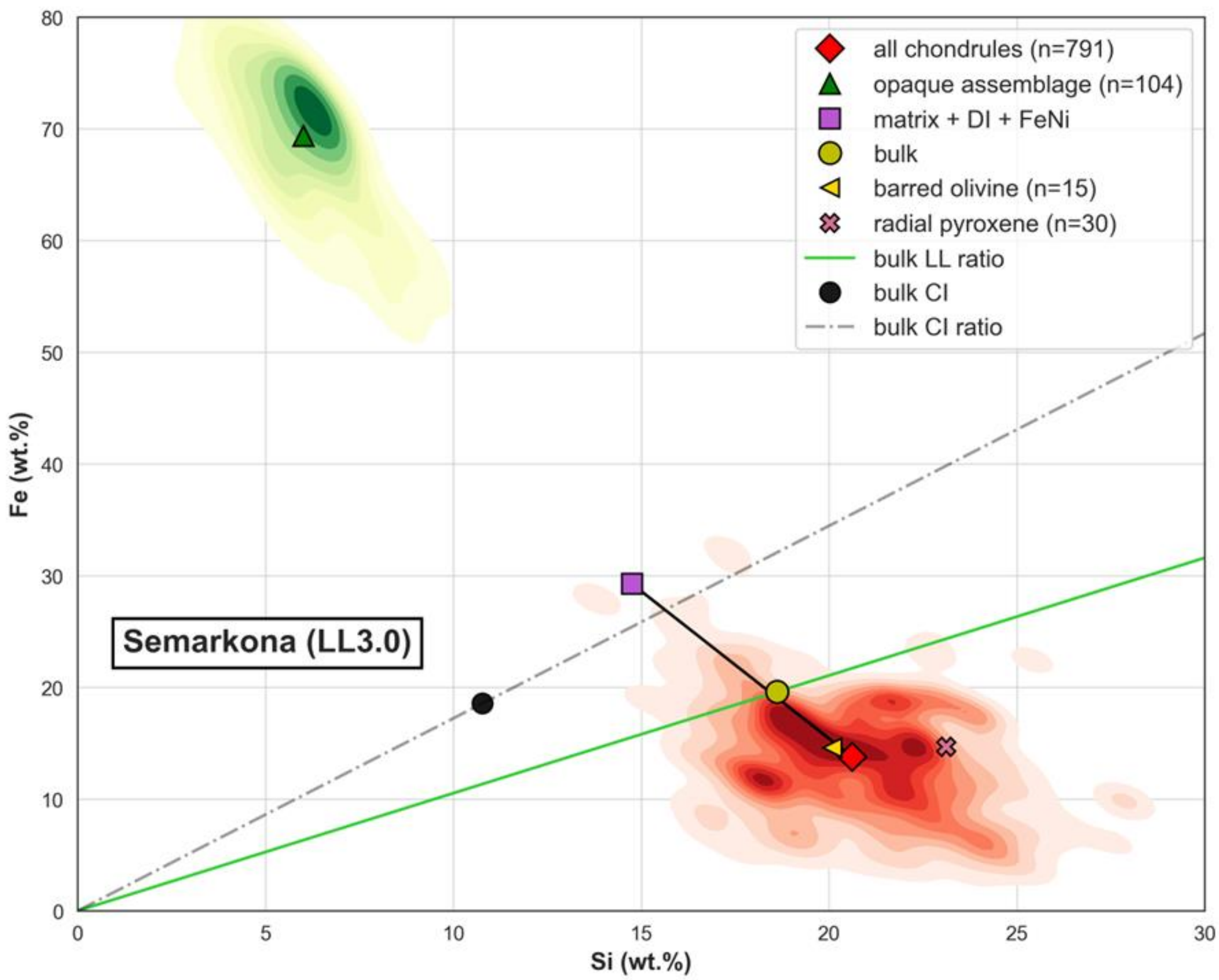

**Figure 8**: Wt% Fe and Si of inclusions and matrix in Semarkona. Weighted density distributions represent the spread of individual chondrules (red) and opaque assemblages (green), with points showing weighted averages for each inclusion type, matrix, and bulk (all pixels). The weighted mean of all 791 chondrules is the red diamond. The CI value is quite different due to LL chondrite depletion in bulk Fe and the dilution of CI mass fractions by volatiles depleted in other chondrite types.

# DISCUSSION

The deviations from bulk meteorite composition observed in the matrix and inclusions for Mg/Si and Fe/Si are of marked and opposing nature. This is true despite the very large spread among inclusion compositions, as is shown by the weighted density distributions in Figs. 5 through 8. In Renazzo (CR2), subchondritic Mg/Si ratios in matrix and superchondritic ratios in chondrules, as well as strong absolute Si concentration differences, provide strong support for the complementary nature of chondrules and matrix, as described by Hezel and Palme (2010) for carbonaceous chondrites.

The fact that experienced meteorite petrographers can often identify the group of a chondrite by inspection implies that separate reservoirs and processes (not addressed here) in those reservoirs in both NC and CC regions, perhaps separated by heliocentric region and time, were involved in the formation of each group. The modal abundances of identifiable clasts, and their sizes, all presented here, indicate that the CR and LL chondrites contain distinct textural, chemical, and sized populations of chondrules, CAIs, and other inclusions sorted into different populations. Our results, combined with earlier work (Jones, 2012; Ebel et al., 2016; Barosch et al., 2020b; Fendrich & Ebel, 2021) provide strong quantitative support for the qualitative observation by Wood (1963) that "different chondrites are markedly dissimilar in the mean size, size distribution, types, and shapes of chondrules they contain, and their chondrule-to-matrix ratio" though "their bulk chemical compositions remain very similar" (cf., Jones, 2012).

The Mg-Si distribution for chondrules in Renazzo (CR2, Fig. 5) reveals a bimodal distribution. This may hint at two populations of chondrules, likely olivine- and pyroxene-rich ones. Marrocchi et al. (2022, 2024) found two FeO-poor (type I) chondrule populations in CR and LL chondrites: a larger-sized, $^{16}O$-poor generation and a smaller-sized, $^{16}O$-rich generation (cf., Clayton et al., 1991). They interpreted the larger chondrules as younger. Although we see a bimodal distribution of chondrules in Mg-Si ratio (Fig. 5) we have not been able to discern any bimodal size distribution that might correlate with an observable chemical or petrological difference (supplemental Figure S5).

## Matrix and macroscopic component abundances

Modal abundances of inclusions and matrix in many chondrite groups have now been measured using x-ray mapping, segmentation, and image analysis methods identical to those employed here. Ornans-type CO, Vigarano-type CV chondrites, and highly primitive Acfer 094 (C2, ungr) were measured by Ebel et al. (2016), Mighei-type CM chondrites were measured by Fendrich and Ebel (2021), and Barosch et al. (2020b) analyzed Kakangari-type K chondrites. Component ratios across all these chondrites, with CI (100% matrix) are presented together in Figure 9.

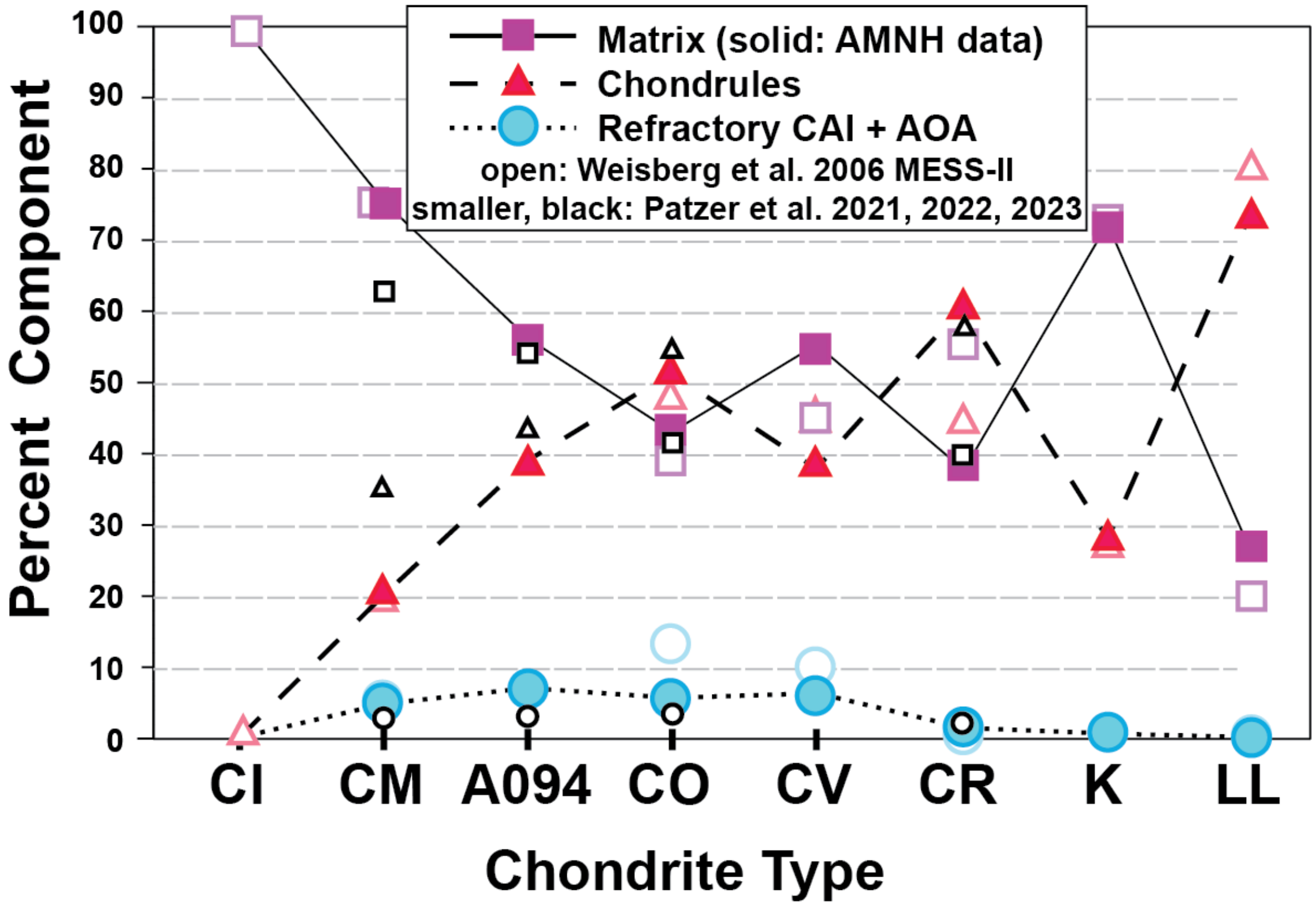

**Figure 9**: Ratios of chondrules, refractory inclusions, and matrix among chondrite groups and primitive ungrouped chondrite Acfer 094, all measured using consistent mapping and segmentation techniques. Literature review (Weisberg et al., 2006) and recent measurements agree well. Patzer et al. (2022) Renazzo data are used rather than their mean CR values. The data in Figure 9 are included as supplementary Table S3.

**Inclusion-matrix complementarity**

Carbonaceous chondrite matrix contains the highly volatile component of those meteorites, including insoluble organic matter and presolar grains that are highly susceptible to oxidation (Huss et al., 2006). These components cannot have experienced the high temperatures implicated in the formation of chondrules, CAIs, and AOAs (Ebel, 2006). Yet substantial evidence supports their coexistence and accretion in single reservoirs for each chondrite group. The paradoxical coexistence of these low-temperature and high-temperature components in chondrites suggests that the high temperature components must have experienced either a highly localized heating process while spatially coexisting with the low-temperature components; or that the components were spatially separated, perhaps vertically in the disk, but within the same chemical "reservoir" of parent body accretion; or that a minor contribution of "CI-like", volatile-rich matrix occurred after chondrule formation (Braukmüller et al., 2018). A single reservoir for chondrule and matrix formation and accretion would provide another important constraint on astrophysical models for these processes.

Complementary distributions of major elements among both clasts and matrix is apparent in this data for CR and LL chondrites. The fact that the present data include vastly more clasts and matrix pixel analyses than previous work lends further evidence to the constraint of matrix and chondrule formation in single reservoirs, or "complementarity", on carbonaceous chondrite formation (Palme 1992; Hezel and Palme 2010). Figures 5 and 6 illustrate that chondrules are superchondritic in Mg/Si and matrix is subchondritic in both CR and LL chondrites. Other carbonaceous chondrites reveal the same relationships between Mg/Si in chondrules relative to matrix (Ebel et al., 2016; Fendrich and Ebel, 2021).

Data presented here demonstrate that the bulk composition of any single inclusion, of any type, yields absolutely no clue as to its host chondrite group, as was demonstrated for other chondrite groups by earlier work (e.g., Hezel & Palme, 2010; Ebel et al., 2016). The composition range among all chondrules and other inclusions is huge. Of course, in these measurements the various components must add up to the bulk composition. It is the additional fact that in carbonaceous chondrite bulk compositions the major element ratios are very nearly CI-chondritic that makes the complementary nature of inclusions and matrix so striking. Similar relationships are seen in CO and CV (Ebel et al., 2016), CM (Fendrich and Ebel, 2021), and K (Barosch et al., 2020b) chondrites.

If inclusions were formed and transported from a separate reservoir from matrix, spatially or temporally, then, given their large variation in composition, the mixing from that reservoir must have been sufficiently non-stochastic that inclusions combined to make exactly the total inclusion composition that complements matrix. Among and within the different carbonaceous chondrite groups, the total inclusion composition must differ so as to exactly balance matrix, because inclusion/matrix ratios vary among and within the chondrite groups (Fig. 9). Such a multiple-reservoir concept violates Ockham's principle of parsimony.

Semarkona (LL3.0) does not meet the entire criteria for "complementarity" proposed by Hezel et al. (2018), which requires a bulk composition close to CI, as quantified by Goldberg et al. (2015). Though bulk Mg/Si of Semarkona differs from that of CI by only 11% (Fig. 1), our results show that Semarkona illustrates the same qualitative Mg/Si relationship shown by the carbonaceous chondrites. This would appear to be a consistent property of both CC and OC meteorites. This consistency is an argument in favor of the hypothesis of formation of LL chondrite matrix and chondrules from a single reservoir, albeit one depleted in Fe and siderophile elements, and Mg relative to Si, compared to CI. While not meeting the criteria for complementarity, sensu stricto, formation of the ordinary chondrites from single reservoir(s) provides another powerful constraint on astrophysical models.

## CONCLUSIONS

We and others (cited) have shown that the compositions of chondrules and matrix, refractory inclusions (AOAs and CAIs), and opaque assemblages differ well beyond analytical error in major element composition. Despite the large differences among carbonaceous chondrite groups in the abundances of the "clastic" components and matrix, all these meteorite groups have very close to solar Mg/Si ratios (Lodders, 2021). Other work has established similar complementary relationships in this and other element ratios (Hezel et al., 2018). The LL chondrites show similar relationships while not meeting the criteria for "complementarity" due to slight depletions in Mg and strong depletions in Fe relative to Si.

The meaning of these observations is clear, and answers the question of MacPherson et al. (2005, sec. 4.4, quoted above). For each chondrite group, the chondrules were formed from a single chemical reservoir that also contained the matrix material, or to which a small fraction of volatile-rich "CI-like" matrix was later added. The chondrite parent bodies accreted from these materials without significant mixing of macroscopic components (chondrules, etc.). The mechanism by which this happened is not revealed by these observations. Our work merely serves to demand the *existence* of a mechanism or mechanisms to result in these observed properties of chondrites.

*Acknowledgments* -- The authors thank AMNH EPS staff over many years for technical help and Drs. H. C. Connolly and J. M. Friedrich for consultation. Parts of this research were supported by U.S. N.A.S.A. grants awarded to DSE and MKW. The AMNH Physical Sciences REU Program (NSF) and NASA grants supported students Isabelle Erb, and (now PhD) co-authors Lobo, Bayron and Gemma in early versions of the work. This research has made use of NASA's Astrophysics Data System Bibliographic Services. Insightful, critical reviews by Dr. Dominik Hezel and an anonymous reviewer are greatly appreciated.

*Editorial Handling* – Yves Marrocchi.

## SUPPLEMENTAL DATA

Supplemental data may be accessed online in a DSpace repository maintained by the American Museum of Natural History Library, https://digitallibrary.amnh.org/home. The complete supplemental data for this publication, at https://doi.org/10.5531/sd.eps.8, is about 380 MB in size.